\documentclass[11pt,a4,twoside,longbibliography]{article}
\usepackage[dvips,a4paper,left=2.0cm,right=2.0cm,top=2.0cm,bottom=2.0cm,headsep=1em]{geometry}
\usepackage{fancyhdr}
\usepackage{titlesec}
\usepackage[latin1]{inputenc}
\usepackage{amsmath,amsfonts,amssymb,array,amsthm}
\usepackage{rotating}
\usepackage[font={small}]{caption}
\usepackage{lmodern,slantsc}
\usepackage{multirow}
\usepackage{chngcntr}
\usepackage{placeins}
\usepackage{caption}
\usepackage{enumitem}
\usepackage{cite}
\usepackage[sort&compress,numbers]{natbib}
\def\bibsep{5pt}
\def\bibfont{\small}
\usepackage[breaklinks,
colorlinks=true, linkcolor=blue, citecolor=blue, urlcolor=blue,
pdfborder={0 0 0}, pdfpagelabels]{hyperref}
\def\vx{\mathbf x}

\def\ve{\mathbf e}

\def\v0{\boldsymbol{0}}

\newlength{\FigureHeight}
\newlength{\FigureHeightHalf}

\numberwithin{equation}{section}
\titleformat{\section}
{\large\bfseries}{\thetitle.}{0.5em}{}
\titleformat{\subsection}
{\normalfont\bfseries}{\thetitle.}{0.5em}{}
\titleformat{\subsubsection}
{\normalfont\itshape}{\normalfont\thetitle.}{0.5em}{}
\begin{document}

\title{\vspace{-2.5em} A critical examination of the conformal invariance\\ in the statistical equations of 2D turbulent scalar fields}

\author{Michael Frewer$\,^1$\thanks{Email address for correspondence:
frewer.science@gmail.com}$\:\,$ \& George Khujadze$\,^2$ \\ \\
\small $^1$ Heidelberg, Germany\\
\small $^2$ Chair of Fluid Mechanics, Universit\"at Siegen, 57068
Siegen, Germany}
\date{{\small\today}}
\clearpage \maketitle \thispagestyle{empty}

\vspace{-2.0em}\begin{abstract}

\noindent
The recent study by Wac{\l}awczyk {\it et al.} [\href{https://doi.org/10.1103/PhysRevFluids.6.084610}{Phys.~Rev.~Fluids~{\bf 6},~084610~(2021)}]
on conformal invariance in 2D turbulence is misleading as it makes three incorrect claims that form the core of their work.\linebreak[4] We will correct these claims and put them into the right perspective: First, the conformal invariance as proposed by Wac{\l}awczyk {\it et al.} is not related to the result that zero-isolines of the scalar field in the inverse energy cascade display a Schramm-Loewner evolution (SLE). Second, the conformal invariance is not a Lie-group for all values of the scalar field since it inherently violates the smoothness axiom of a Lie-group action, with the effect that a physical PDF gets mapped to a non-physical one. Third, although Wac{\l}awczyk {\it et al.} recognize that their conformal invariance does not constitute a symmetry but only a weaker equivalence transformation, it is still not classified correctly. The~claim that their equivalence can map between solutions is not true. This fact will be demonstrated by using an illustrative example.

\vspace{0.5em}\noindent{\footnotesize{\bf Keywords:} {\it Statistical Physics, Conformal Invariance, 2D Turbulence, Symmetry Analysis, Lie-Groups,\\ Probability Density Functions,
Stochastic Differential Equations, Closure Problem}}
\end{abstract}

\vspace{0em}
\section{A brief summary of the key results in \cite{Grebenev21}\label{Sec1}}

Without providing a derivation of its origin, the following new conformal transformation is proposed (Eqs.$\,$(25)-(32),(48)~in~\cite{Grebenev21})
\begin{equation}
\left.
\begin{aligned}
\!\!\!\!&t^*=t,\;\; \vx^*=\boldsymbol{\mathcal{X}}(\vx,a),\;\;
\vx^{\prime *}=\boldsymbol{\mathcal{X}}(\vx,a)+\gamma^{1/3}(\vx)\Big[(\vx^\prime-\vx)\cdot\nabla\Big]\boldsymbol{\mathcal{X}}(\vx,a),\\[0.5em]
\!\!\!\!&\phi^*=\gamma^{-1}(\vx)\,\phi,\;\; \phi^{\prime *}=\gamma^{-m/6}(\vx)\,\phi^\prime,\;\;
f_1^*=\gamma(\vx)\,f_1,\;\; f_2^*=\gamma^{(6+m)/6}(\vx)\,f_2,\;\; Q^*=\gamma^{-2}(\vx)\,Q,
\end{aligned}
~~~\right\}
\label{211023:1700}
\end{equation}
with the local scaling factor
\begin{equation}
\gamma^{-1}(\vx)\,=\,\big(\partial_x \mathcal{X}\big)^2+\big(\partial_y \mathcal{X}\big)^2\,=\,\big(\partial_x \mathcal{Y}\big)^2+\big(\partial_y \mathcal{Y}\big)^2,
\end{equation}
where $\boldsymbol{\mathcal{X}}=(\mathcal{X},\mathcal{Y})$ forms a 2D vector pair of harmonic conjugates $\mathcal{X}$ and $\mathcal{Y}$ parametrized by~$a$.
 As explicitly shown in \cite{Grebenev21}, it indeed leaves invariant the unclosed PDF-equation (Eqs.$\,$(5)-(8)~in~\cite{Grebenev21})
\begin{equation}
\partial_t f_1+\nabla\cdot\!\int\! d\vx^\prime d\phi^\prime\phi^\prime\ve_z\times\frac{\vx-\vx^\prime}{\vert\vx-\vx^\prime\vert^m}f_2=\alpha\,\partial_\phi(\phi f_1)+Q\,\partial^2_\phi f_1,
\label{211019:0005}
\end{equation}
but only if its additionally evaluated in the end on the scalar zero-isoline $\phi^*=\phi=0$ {\it after} the transformation~\eqref{211023:1700} has been applied. Equation \eqref{211019:0005} is a truncated and thus dynamically unclosed equation for the 1-point probability density function (PDF) $f_1$, representing only the first equation in an infinitely coupled chain of equations for the multi-point PDFs $f_n$. It samples through a class of 2D hydrodynamic models (set by the exponent $m$) for a scalar field $\phi(t,\vx)$, subjected to zero small-scale viscosity ($\nu=0$), large-scale friction $\alpha\geq 0$ and stochastic forcing $Q$. Specifically for $m=2$ the scalar~$\phi$ represents the vorticity $\omega$ of 2D Navier-Stokes turbulence up to a re-scaled model constant $\beta$ (Eqs.$\,$(3)-(4) in \cite{Grebenev21}). The dependencies in \eqref{211019:0005}~are: $f_1=f_1(t,\vx,\phi)$, $f_2=f_2(t,\vx,\phi,\vx^\prime,\phi^\prime)$ and~$Q=Q(\vx)$.

\noindent
For $Q=0$ (zero stochastic forcing), the unclosed equation \eqref{211019:0005} turns into a first-order equation of hyperbolic type of the form (when identifying the 2-point PDF as $f_2=(f_2/f_1)f_1$)
\begin{equation}
F(x_0,x_1,x_2,x_3,u,p_0,p_1,p_2,p_3)=0,
\end{equation}
where $(x_0,x_1,x_2,x_3)=(t,x,y,\phi)$, $u=f_1$, and $(p_0,p_1,p_2,p_3)=(\partial_t f_1,\partial_x f_1,\partial_y f_1,\partial_\phi f_1)$, allowing it to now formulate equation \eqref{211019:0005} as the following unclosed set of characteristic equations (yielding the so-called Monge curves, see e.g. \cite{Courant62})
\begin{equation}
\frac{dx_i}{ds}=\partial_{p_i}F,\quad\;\; \frac{du}{ds}=p_i\partial_{p_i}F,\quad\;\; \frac{dp_i}{ds}=-(p_i\partial_u+\partial_{x_i})F, \quad\;\; i=0,\dotsc,3,
\label{211023:2132}
\end{equation}
for the $(2\cdot 4+1)=9$ functions $x_i$, $u$, $p_i$ of a parameter $s$, each supplemented with an own independent initial condition at $s=0$. In \cite{Grebenev19} for $m=2$, $\alpha=0$, and in \cite{Grebenev20} for arbitrary $m>1$, $\alpha\geq0$, it has been shown that if the defining equation~\eqref{211019:0005} is conformally invariant, then so are the first two sets of characteristic equations in \eqref{211023:2132}, which take here the explicit form \cite{Grebenev19,Grebenev20}
\begin{equation}
\left.
\begin{gathered}
\frac{dt(s)}{ds}=1,\quad\;\;
\frac{d\vx(s)}{ds}=\int\! d\vx^\prime d\phi^\prime\phi^\prime\ve_z\times\frac{\vx(s)-\vx^\prime}{\vert\vx(s)-\vx^\prime\vert^m}
\frac{f_2(s,\vx^\prime,\phi^\prime)}{f_1(s)},\quad\;\;
\frac{d\phi(s)}{ds}=-\alpha\cdot \phi(s),\\[0.5em]
\frac{df_1(s)}{ds}=-f_1(s)\,\nabla\Big\vert_{\vx=\vx(s)}\cdot\!\int\! d\vx^\prime d\phi^\prime\phi^\prime\ve_z\times\frac{\vx-\vx^\prime}{\vert\vx-\vx^\prime\vert^m}
\frac{f_2\big(t(s),\vx,\phi(s),\vx^\prime,\phi^\prime\big)}{f_1\big(t(s),\vx,\phi(s)\big)}+\alpha f_1(s),
\end{gathered}
~~~~~\right\}
\label{211023:2217}
\end{equation}
where $f_1(s):=f_1\big(t(s),\vx(s),\phi(s)\big)$ and $f_2(s,\vx^\prime,\phi^\prime):=f_2\big(t(s),\vx(s),\phi(s),\vx^\prime,\phi^\prime\big)$. In both \cite{Grebenev19} and \cite{Grebenev20}, however, the third set of characteristic equations in \eqref{211023:2132} are not included and therefore are missing in their analysis for unknown reasons. Hence, their conformal invariance analysis on the characteristic equations of the defining equation~\eqref{211019:0005} is incomplete, a remark that also applies to \cite{Grebenev21} when claiming {\it ``that the zero-scalar characteristics of the equations are conformally invariant"} [p.$\,$2].

As discussed in \cite{Friedrich12.1,Friedrich12.2}, it is tempting to interpret the characteristic equations \eqref{211023:2132} and their higher orders in the infinite chain as a kind of Lagrangian dynamics of quasi-particles moving in an averaged field. However, it is not to be confused with the statistics of the true Lagrangian framework, as it seems to have happened in \cite{Grebenev19,Grebenev20} and again in \cite{Grebenev21} by claiming that in their earlier studies {\it ``the CG [conformal group] invariance both for the Lagrangian path and the 1-point PDF of vorticity, i.e., $f_1(x,\omega,t)$ taken along the zero-vorticity characteristics was established"}~[p.$\,$2].

Fact is, the statistics of a Lagrangian trajectory tracking a fluid particle follows different dynamical rules than those characteristic equations \eqref{211023:2132} obtained from an Eulerian description. For a derivation of the PDF equations in the Lagrangian description, see e.g. \cite{Friedrich02}. It are those equations and its characteristics which describe the statistics of a Lagrangian path, but since they are not part of the analysis in \cite{Grebenev19,Grebenev20,Grebenev21}, nothing can be said about the statistical invariance properties of the infinite chain of PDF equations sampling a Lagrangian path in 2D turbulence.

\section{The non-relation to Schramm-Loewner evolution (SLE)\label{Sec2}}

On the one side there is the conformal transformation \eqref{211023:1700} proposed by Wac{\l}awczyk {\it et al.}, which leaves invariant the unclosed PDF equation \eqref{211019:0005} and part of its characteristic equations  \eqref{211023:2217}, provided that the obtained transformed equations are additionally evaluated on the zero-isoline of the scalar field~$\phi$. This invariance is about an invariance admitted by an unclosed statistical {\it equation} and not about an invariance that maps between {\it solutions} --- see Sec.$\,$\ref{Sec4} for more details on this fact, where it will be explicitly shown what it ultimately means for the conformal invariance~\eqref{211023:1700} not to be a symmetry, but only an equivalence transformation of an unclosed equation, unable to explore the space of physically realizable solutions. Also important to note is that when putting $\phi=0$ to achieve conformal invariance in the unclosed equations does not mean that $\phi=0$ itself stays conformally invariant. For this, the invariant solution concept must be invoked, but which is not part of the analysis in \cite{Grebenev21} and also cannot be as long as unclosed equations are considered.

On the other side there is the numerical result, as first initiated by Bernard {\it et al.} \cite{Bernard06} for the specific case $m=2$ and later generalized for arbitrary $m<4$ in
\cite{Falkovich07,Bernard07,Falkovich10}, which states that when {\it solving} the underlying stochastic equation defining \eqref{211019:0005} at any fixed time $t$ in the statistically stationary and inverse cascade regime, the zero-isoline of the scalar field $\phi=\phi(t,\vx)=0$ traces out a 2D random curve that can be conformally mapped to a 1D Brownian walk. To achieve a statistically stationary 2D flow in the inverse energy cascade regime, the following stochastic equation was considered, for example, in \cite{Falkovich10}
\begin{equation}
\partial_t \phi(t,\vx)-\nabla\phi(t,\vx)\cdot\!\int\! d\vx^\prime\phi(t,\vx^\prime)\,\ve_z\times\frac{\vx-\vx^\prime}{\vert\vx-\vx^\prime\vert^m}=-\nu_H\Delta^8\phi(t,\vx)-\alpha\phi(t,\vx)
+L(t,\vx),
\label{211024:2252}
\end{equation}
where $L$ is Gaussian white noise with zero mean and correlator $\langle L(t,\vx)L(t^\prime,\vx^\prime)\rangle=2Q(\vx,\vx^\prime)\delta(t-t^\prime)$ and $\nu_H>0$ a higher-order viscosity. The three coefficients, the forcing Q, friction $\alpha$ and viscosity~$\nu_H$, are now to be adjusted in such a way that a well-formed stationary inverse energy cascade can develop. The task of the stochastic forcing $L$, peaked around a wavenumber $k_f\sim100$, is to pump energy not at large scales but at small scales into the system to sustain the scalar fluctuations. The reason is to be able to halt the simultaneously developing direct enstrophy cascade at wavenumbers $k>k_f$ by means of the introduced hyper-viscous damping $\nu_H\ll 1$ to extend the inertial range of the inverse cascade,\linebreak[4] while the friction term $\alpha\sim 1$ removes energy at large scales, thereby making the inverse energy cascade then stationary. Hence, only in such a fine-tuned system with a well-developed inertial range of the inverse energy cascade, as shown in~\cite{Bernard06,Falkovich07,Bernard07,Falkovich10}, will the obtained large-scale solution of \eqref{211024:2252} on a zero-isoline conformally map to a 1D Brownian walk. In other words, the spatially large-scale solutions of \eqref{211024:2252} resulting from a steady-state inverse energy cascade on a zero-isoline belong to the class of Schramm-Loewner evolution ($\text{SLE}_\kappa$) curves, characterized by the respective dimensionless diffusivity~$\kappa$. In \cite{Falkovich10} it is conjectured that all solutions in the range $3\leq m<4$ belong to $\kappa=4$, while those in $0<m\leq 3$ to $\kappa={12/m}$. Notice the different defining range of the hydrodynamical-model parameter $m$ used in \cite{Falkovich10} and \cite{Grebenev21}, where the range of $m$ as used herein is that of \cite{Grebenev21} and relates to $4-m$ as used in \cite{Falkovich10}. To note is also, as shown in \cite{Falkovich10}, that an inverse energy cascade can only exist for models $(4-m)>0$, i.e., when specifying $m>4$ no inverse energy cascade will develop.

Comparing now the result obtained by Wac{\l}awczyk {\it et al.} \cite{Grebenev21} with that of Bernard and Falkovich {\it et al.} \cite{Bernard06,Falkovich07,Bernard07,Falkovich10}, it is clear that no relation or link exists, nor can one be established.

\textbf{a. Unclosed equations~vs.~particular solutions:}
The study \cite{Grebenev21} is about a conformal\linebreak[4] equivalence being admitted by unclosed statistical equations, while the studies \cite{Bernard06,Falkovich07,Bernard07,Falkovich10} are about how a particular solution of a large-scale zero-isoline of the scalar field can be conformally mapped to a Brownian walk. That is, the latter is about the growth of a random fractal curve where each incremental step is produced by a conformal transformation characterized by a Brownian motion, and not about some conformal equivalence \eqref{211023:1700} that just leaves invariant an unclosed statistical equation~\eqref{211019:0005}.

\textbf{b. Analytically global~vs.~numerically local:}
The conformal map determined or used\linebreak[4] in~\cite{Bernard06,Falkovich07,Bernard07,Falkovich10} for SLE is always known only locally and never globally, in clear contrast to the conformal map in \cite{Grebenev21} which can be expressed in the analytically exact closed form \eqref{211023:1700}. The SLE conformal map $g_s(z)$ in \cite{Bernard06,Falkovich07,Bernard07,Falkovich10} follows the rule of the Loewner equation
\begin{equation}
\frac{\partial g_s(z)}{\partial s}=\frac{2}{g_s(z)-\xi(s)}, \quad g_0(z)=z,
\end{equation}
an initial value problem driven by a Brownian process $\xi$ for which no exact solution for the mapping $g_s$ is known~yet. Nevertheless, as it was done in \cite{Bernard06,Falkovich07,Bernard07,Falkovich10}, the above equation can be solved numerically such that a 2D random fractal curve $\Gamma(s)$, if properly parametrized by $s$, gets locally and conformally mapped by $g_s$ to $\xi(s)$, i.e., where the driving function $\xi$ and the curve~$\Gamma$ are then related by\linebreak[4] $\xi(s)=g_s(\Gamma(s))$ within the numerical precision considered. Hence, no link between \cite{Grebenev21} and \cite{Bernard06,Falkovich07,Bernard07,Falkovich10} can be established, in particular as the latter considers full random fractal quantities to be transformed, while the former only averaged or statistically sampled quantities.

\textbf{c. Arbitrary system~vs.~fine-tuned system:}
The conformal invariance \eqref{211023:1700} of \eqref{211019:0005} is valid for {\it any} configuration of the parameters $m$, $\alpha$ and $Q$. That is, irrespective of the hydrodynamic model $m>1$ chosen, with or without a large-scale friction or a stochastic forcing, and also irrespective of whether such a forcing will inject energy at large or small scales into the system, in all cases this conformal equivalence will be admitted. Only when including a small-scale viscous force it will break,\pagebreak[4] 

\newgeometry{left=2.0cm,right=2.0cm,top=2.0cm,bottom=1.5cm,headsep=1em}

\noindent
as shown in \cite{Grebenev20,Grebenev21}. Hence, irrespective of whether the flow will be statistically stationary or not, or whether an inverse energy cascade will develop or not, the conformal invariance in~\cite{Grebenev21} will always exist. This is of course in clear contrast to the result obtained in \cite{Bernard06,Falkovich07,Bernard07,Falkovich10}, where the conformal map of SLE, as described before, is only obtained in a highly fine-tuned system. A balanced interplay between $m$, $\alpha$ and~$Q$ along with a hyper-viscous damping $\nu_H$ is needed in order to yield a well-developed inertial range of the inverse energy cascade, the only regime in 2D turbulence where conformal invariance has been seen so far.

Hence, there are enough clear signs that there is no link between the findings of \cite{Bernard06,Falkovich07,Bernard07,Falkovich10} and the result obtained in~\cite{Grebenev21}, as misleadingly claimed therein by saying: {\it ``To link the results obtained in Refs.~[11,13,14] with the results of Refs.~[3,7], we presently consider the first equation from the LMN chain for 2$d$ scalar fields $\phi$ under Gaussian white-in-time forcing and large-scale friction"} [p.$\,$2]. It's not enough to just include an arbitrary stochastic Gaussian force and a large-scale friction into the system in order to make such a link. As already said, much more is needed, let alone the fact that a hyper-viscous term, which is a necessary ingredient in \cite{Bernard06,Falkovich07,Bernard07,Falkovich10}, is missing in \cite{Grebenev21} for a reason, as it would only break the conformal equivalence \eqref{211023:1700}.

\section{Conformal invariance not the Lie-group from \cite{Grebenev19,Grebenev20}\label{Sec3}}

The transformation \eqref{211023:1700} is a new proposal by Wac{\l}awczyk {\it et al.} for a conformal equivalence. It is not the full global form of the original Lie-group infinitesimals determined in their earlier studies \cite{Grebenev17,Grebenev19,Grebenev20} and refuted as a conformal Lie-group in \cite{Frewer18,Frewer21.1,Frewer21.2,Frewer21.3}. The decisive difference is that \eqref{211023:1700} is missing the infinitesimal Lie-group constraint for the scalar field $\xi^\phi_x=\xi^\phi_y=0$, which translates to the global constraint $\partial_x\gamma(\vx)=\partial_y\gamma(\vx)=0$. A consistent Lie-group invariance analysis shows that this constraint has to hold for {\it all} isolines of $\phi$, including the one for $\phi=0$, thus breaking the conformal group \cite{Frewer18,Frewer21.1,Frewer21.2,Frewer21.3}.\linebreak[4] On the other hand, this constraint can surely be switched off for $\phi=0$, but it then comes at the price of leading to internal inconsistencies in the Lie-group invariance analysis, as clearly demonstrated\linebreak[4] in~\cite{Frewer18,Frewer21.1,Frewer21.2,Frewer21.3}. Only when breaking the conformal group for {\it all} scalar isolines an overall consistent analysis is restored.

When ignoring these inconsistencies, which can and has been done in \cite{Grebenev21} by proposing \eqref{211023:1700}, it is not surprising that it has consequences for all later implications. A Lie-group analysis does not give this warning without reason. For example, one of the problems of \eqref{211023:1700} is that since it's only admitted as a conformal equivalence for $\phi=0$, while for all other values $\phi\neq 0$ only if $\gamma$ is a spatial
constant ($\partial_x\gamma(\vx)=\partial_y\gamma(\vx)=0$), it inevitably leads to a nonphysical discontinuity (jump) in the transformation for the PDFs $f_1$ and $f_2$ at $\phi=0$. And this discontinuity is present irrespective of the flow configuration and the flow regime considered. Imagine the inertial regime of the inverse energy cascade at a fixed point in space, then a discontinuously mapped PDF at $\phi^*=\phi=0$ will mean that when statistically sampling the scalar field $\phi^*$ at $0$ it will be different than when sampling it at $\lim_{\epsilon\to 0}(0+\epsilon)$, which clearly is a nonphysical behaviour of a PDF in fully developed turbulence.
Such non-physical discontinuous PDFs can be used as initial conditions, but they will not develop as such during long time evolution.

In this context note that the realizability check of transformation \eqref{211023:1700} in Sec.$\,$V in \cite{Grebenev21} is misleading and meaningless. The check is only performed for the discrete value $\phi=0$, although it~should have been done for all values $\phi$, at least in some infinitesimal region around $\phi=0$. This would have explicitly revealed the meaninglessness of the mapped PDFs. It is clear that the example in Sec.$\,$V~in~\cite{Grebenev21} wants to demonstrate how a hypothetical global-solution (for all $\phi$) can be conformally mapped to a single point-solution (only for $\phi=0$). But, it should also be clear that it's a mapping between PDFs~$f_n$, which are entities that define a physical probability measure $f_n d\phi$ over a range $d\phi$ and not over a single point (having zero probability). And since there is no solution map beyond the point $\phi=0$~(except~for~constant~$\gamma$), there is no use in a PDF if no solution probability can~be~determined.

\section{Conformal invariance not mapping between solutions\label{Sec4}}

On p.~5 in \cite{Grebenev21} it is said: {\it ``This scaling invariance has some important implications. For example,\linebreak[4] once the solution of equation~(5) in the original variables is known, the statistics of the rescaled field and\hfill $\partial f_1^*/\partial t^*$\hfill are\hfill also\hfill determined".}\hfill This\hfill statement\hfill is\hfill indeed\hfill true\hfill for\hfill the\hfill classical\hfill scaling\hfill invariance\pagebreak[4]

%\restoregeometry
\newgeometry{left=2.0cm,right=2.0cm,top=2.0cm,bottom=2.0cm,headsep=1em}

\noindent
considered in Sec.$\,$III by Eqs.$\,$(18)-(20) in \cite{Grebenev21}, but it's not true anymore for their conformal invariance Eqs.$\,$(25)-(32),(48) considered in Sec.$\,$IV, claiming that with this invariance a {\it ``solution of equation~(5) for a homogeneous field could determine solution for an inhomogeneous~one"} --- see also the explanation to equation (8) in \cite{Grebenev21.1} where this incorrect assertion is even stressed more forcefully and without residual doubt.

The reason for this different inference of those two scalings is that the classical scaling is an equivalence that leaves invariant the defining and closed stochastic Eq.$\,$(14) in \cite{Grebenev21}, which then in turn uniquely implies the corresponding invariance for {\it all} resulting statistical equations, while their conformal invariance, however, is clearly opposite to that. It's an equivalence that is only admitted by a particular unclosed statistical equation and {\it not} by its defining stochastic Eq.$\,$(14) anymore. That is, their conformal invariance is an equivalence transformation that only maps between unclosed equations and no longer between solutions anymore, as it's the case with the classical invariance. Keep in mind that the conformal invariance in Eq.$\,$(5) only arises because of exploiting the fact that the unclosed (non-modelled) 2-point PDF $f_2$ is identified as an own independent variable to transform independently of the 1-point PDF $f_1$, i.e., the conformal invariance transforms one equation with two independent functions, a~degree of freedom that is not present in the underlying and defining stochastic Eq.$\,$(14) and therefore, obviously, does not allow for such an invariance. Hence, for all statistical invariances where there is no relation to its defining stochastic equation, extra caution has to be exercised in whether such invariances are realizable or not.

Although Wac{\l}awczyk {\it et al.} recognize their conformal invariance as an equivalence in Sec.$\,$IV.A, it is not recognized to its full extent. They identify it as an equivalence only because of having to transform the parameter~$Q$, but not for the far more important reason that the considered equation itself is unclosed. Transforming parameter values to new values is not the problem since this indeed may allow to map solutions to new solutions, as it's the case for the classical scaling in Sec.$\,$III in \cite{Grebenev21}. The problem is mapping unclosed equations to new unclosed equations, where solutions in general are no longer mapped to solutions anymore, as it's clearly the case for their conformal transformation in Sec.$\,$IV, in particular as it also violates an essential solution constraint (Eq.$\,$(11) in \cite{Grebenev21}) on top of that.

To render our demonstration as simple as possible, we consider a model that in principle mimics the set-up and result in \cite{Grebenev21}. The starting point in \cite{Grebenev21} is Eq.$\,$(14), a stochastically forced nonlinear dynamical equation that can attain a statistically stationary regime. To mimic these characteristics in principle, we consider the following nonlinear stochastic ODE
\begin{equation}
\frac{d{\hat{\phi}}}{dt}+\rho\hat{\phi}^3=\hat{\alpha}\hat{\phi}+\hat{L}(t),\quad \rho, \hat{\alpha}>0,
\label{211030:1406}
\end{equation}
similarly structured as Eqs.$\,$(14), with the only difference that instead of a nonlocal nonlinearity $\hat{\phi}\!\int\! dt^\prime \hat{\phi}^\prime K(t,t^\prime)$ we have chosen a local cubic nonlinearity (overdamped motion in a stable quartic potential). We made this choice just for the sake of simplicity, in order to not only obtain stable random solutions but also an overall steady-state PDF~with converging moments, which, for example, for a quadratic nonlinearity (cubic potential) is impossible to obtain throughout the whole stationary regime due to a not sufficiently fast decaying PDF \cite{Ryabov19}. The stochastic force $\hat{L}$ in \eqref{211030:1406} is again a white noise with zero mean and correlator $\langle \hat{L}(t)\hat{L}(t^\prime)\rangle= 2\hat{Q}\delta(t-t^\prime)$, defining in turn a Wiener process
$dw/dt=\hat{L}$ with $\langle w(t)w(t^\prime)\rangle= 2\hat{Q}\min(t,t^\prime)$. Taking now the same scaling for $\hat{\phi}$ as for $\phi$ in \cite{Grebenev21}, the following particular scaling invariance is admitted by \eqref{211030:1406}
\begin{equation}
t^*=\lambda^2 t,\quad
\hat{\phi}^*=\lambda^{m-4}\hat{\phi},\quad
\rho^*=\lambda^{-2m+6}\rho,\quad
\hat{\alpha}^*=\lambda^{-2}\hat{\alpha},\quad
\hat{L}^*=\lambda^{m-6}\hat{L},\quad
\hat{Q}^*=\lambda^{2m-10}\hat{Q},
\label{211101:1810}
\end{equation}
which by construction matches the scaling Eqs.$\,$(18)~and~(20)~in \cite{Grebenev21}, up to the transformation of the newly introduced parameter $\rho$. When considering the statistics of \eqref{211030:1406}, it can be described, for example, as in \cite{Grebenev21}, by an infinite hierarchy of PDF equations, here though for multi-time. However, since in~\eqref{211030:1406} we have chosen a local nonlinearity instead of a nonlocal one, the multi-time hierarchy decouples when applying the derivation procedure of \cite{Friedrich12.1}, where the first equation then reduces to the closed PDF equation
\begin{equation}
\partial_t f=-\partial_{\hat{\phi}}\big[\big(\hat{\alpha}\hat{\phi}-\rho\hat{\phi}^3\big)f\big]+\hat{Q}\partial^2_{\hat{\phi}}f,
\label{211101:1719}
\end{equation}
being identical to the associated Fokker-Planck equation of the stochastic process \eqref{211030:1406}. Now, since
the\hfill stochastic\hfill equation\hfill \eqref{211030:1406}\hfill defines\hfill and\hfill determines\hfill {\it all}\hfill
statistical\hfill relations\hfill and\hfill not\hfill opposite,\hfill its\hfill

\restoregeometry

\noindent
invariance~\eqref{211101:1810} thus uniquely transfers to \eqref{211101:1719}. Due to the defining relation $f=\langle\delta(\hat{\phi}-\hat{\phi}(t))\rangle$ \cite{Friedrich12.1}, the PDF transforms as
\begin{equation}
f^*=\frac{1}{\lambda^{m-4}}f,
\label{211101:2005}
\end{equation}
leaving then invariant equation \eqref{211101:1719} and its normalization $\int\! d\hat{\phi} f=1$. But also all other statistical relations that follow from \eqref{211030:1406} automatically stay invariant under \eqref{211101:1810}, as for example the infinite hierarchy of the stationary moment equations, where the first two read \cite{Gihman72}
\begin{equation}
\hat{\alpha}\langle\hat{\phi}\rangle-\rho\langle\hat{\phi}^3\rangle=0, \qquad \hat{\alpha}\langle\hat{\phi}^2\rangle-\rho\langle\hat{\phi}^4\rangle+\hat{Q}=0.
\label{211101:1853}
\end{equation}
The above moments can be retrieved analytically through the steady-state solution $f_s$ of \eqref{211101:1719}
\begin{equation}
\langle\hat{\phi}^n\rangle =\!\int\! d\hat{\phi}\, \hat{\phi}^n f_s, \qquad f_s=c\cdot(\rho/\hat{\alpha})^{1/2}e^{-\frac{\hat{\phi}^2}{2\hat{Q}}\big(\frac{\rho}{2}\hat{\phi}^2-\hat{\alpha}\big)},
\label{211108:1044}
\end{equation}
where $c$ is a dimensionless normalization constant such that $\int\! d\hat{\phi} f_s=1$. Notice that since this constant\linebreak[4] transforms invariantly, the above steady-state solution $f_s$ constitutes an invariant solution under the scaling~\eqref{211101:1810} and~\eqref{211101:2005}.

To explicitly see now that the equivalence \eqref{211101:1810} maps statistical solutions to new statistical solutions, let's consider the configuration: $\rho=\hat{\alpha}=1$ and $\hat{Q}=1/8$. Solving the stochastic equation \eqref{211030:1406} numerically for a particular initial condition, say $\hat{\phi}(0)=1$, and then by taking for $T\gg\tau$ the time average $\langle F(\hat{\phi})\rangle=\int_\tau^T dt F(\hat{\phi})/(T-\tau)$ when the motion settles in the statistically stationary regime $t>\tau$, we obtain for the first four moments (up to 2 decimals accurate)
\begin{equation}
\langle\hat{\phi}\rangle=\langle\hat{\phi}^3\rangle=0, \quad \langle\hat{\phi}^2\rangle=0.85, \quad \langle\hat{\phi}^4\rangle=0.98.
\end{equation}
Transforming this result then by the above invariance, e.g. for $\lambda=1/2$ and $m=2$, we obtain the transformed values
\begin{equation}
\langle\hat{\phi}^*\rangle=\langle\hat{\phi}^{*3}\rangle=0, \quad \langle\hat{\phi}^{*2}\rangle=0.85\cdot \lambda^{2m-8}, \quad \langle\hat{\phi}^{*4}\rangle=0.98\cdot\lambda^{4m-16},
\label{211102:0010}
\end{equation}
which indeed are the new solutions of the transformed configuration: $\rho^*=1/4$, $\hat{\alpha}^*=4$ and $\hat{Q}^*=8$.

Now let's continue as in \cite{Grebenev21} when in Sec.$\,$IV the conformal equivalence gets introduced as an equivalence only admitted by an unclosed statistical equation and not by it's underlying defining stochastic equation. Here such a procedure would be equivalent by performing an invariance analysis only on the unclosed moment equations \eqref{211101:1853} while ignoring the existence of the defining equation~\eqref{211030:1406}. Since the odd-moment equation in \eqref{211101:1853} evaluates identically to zero, the procedure reduces here to an invariance analysis of the even-moment equation only. Exploiting the fact that it's one equation with two unknowns, we can use this additional degree of freedom to propose a whole new class of equivalence transformations that leaves invariant the even-moment equation in \eqref{211101:1853}, for example, by choosing the
following one:
\begin{equation}
\langle\hat{\phi}^{2}\rangle^*=e^{a_s}\langle\hat{\phi}^{2}\rangle, \quad \langle\hat{\phi}^{4}\rangle^*=e^{a_s}\langle\hat{\phi}^{4}\rangle,
\quad \hat{Q}^*=e^{a_s}\hat{Q},
\quad \rho^*=\rho,
\quad \hat{\alpha}^*=\hat{\alpha},
\label{211102:0004}
\end{equation}
which Wac{\l}awczyk {\it et al.} refer to as the intermittency symmetry \cite{Oberlack14,Oberlack17}.\footnote{In the appendix we show how \eqref{211102:0004} transfers to the entire infinite chain of moment equations, giving by construction an identical scaling for all moments in the leading order. A second and a third alternative ``intermittency symmetry" will also be presented. Subsequently, the nonrealizability for such kind of  transformations is discussed in detail.}
But this invariance is not realizable, i.e., the transformed moments cannot be realized by the defining stochastic
equation \eqref{211030:1406}, no matter which starting configuration in the parameters or which mapping in the variables is chosen. The problem trivially lies in the identical scaling of the second and fourth moment, which thus makes it a nonphysical invariance \cite{Frewer15,Frewer16,Frewer17,Frewer14}. Essentially it violates the classical principle of cause and effect, since no cause in \eqref{211030:1406} exists such that the invariance \eqref{211102:0004} can result as an effect, where it's important to note here that the cause itself need not to be an invariance in order to induce an invariance as an effect \cite{Frewer15,Frewer16,Frewer17,Frewer14}. Hence, the equivalence \eqref{211102:0004} does not map solutions to solutions anymore, in clear contrast to \eqref{211102:0010} that results from \eqref{211101:1810}.

\noindent
For a general configuration $m$, the conformal invariance in \cite{Grebenev21} faces the same methodological problem as the above nonphysical equivalence \eqref{211102:0004}, in that solutions do not get mapped to solutions for the very same reason that it violates the principle of cause and effect. No transformation of the defining stochastic equation for general $m$ exists, whether as an invariance or not, such that the proposed conformal transformation results as a statistical invariance of the considered PDF equation. The only exception where such a causal transformation exists is when the configuration is set to $m=6$. Only then a cause-effect relation can be established, because $m=6$ is the only case in which a consistent mapping both for the two scalar fields $(\phi,\phi^\prime)$ and the two PDF solutions $(f_1,f_2)$ is obtained. For all other values of $m\neq 6$, however, not.

For the scalar fields Eqs.$\,$(28)-(29) it is easily seen that for general $m$ the mapping turns inconsistent when $\vx^\prime=\vx$, except for $m=6$. The reason is simple: since $\vx^\prime=\vx$ implies $\phi^\prime=\phi$ by definition, and since $\vx^{\prime *}=\vx^*$ results from Eqs.$\,$(25)-(27) when $\vx^\prime=\vx$, we must have $\phi^{\prime *}=\phi^*$ when $\vx^\prime=\vx$, but this can only be realized when $m=6$. It is also for this reason that the conformal invariance, except for $m=6$, is inconsistent to the coincidence constraint (Eq.$\,$(11) in \cite{Grebenev21}), an internal PDF constraint which is an essential ingredient when deriving the hierarchy of PDF equations. To note is that the consistency proof in \cite{Grebenev19}, showing that the coincidence constraint stays invariant under the conformal mapping, is flawed. The error in their proof is that only specially designed invariant test functions are used to show their case. But in order to draw the correct conclusion, the invariance must be proven for {\it all} test functions, which clearly fails for $m\neq 6$.

For the PDF functions Eqs.$\,$(30)-(31) it's also easy to see that the mapping for $m\neq6$ is inconsistent and thus nonphysical when trying to construct a cause-effect relationship, however, for a different reason than for the scalar fields. When looking at the definitions of the PDFs $f_1=\langle\delta(\phi_1-\phi(\vx_1,t))\rangle$ and $f_2=\langle\delta(\phi_1-\phi(\vx_1,t))\delta(\phi_2-\phi(\vx_2,t))\rangle$~\cite{Friedrich12.1}, it is easy to see that if one tries to construct a cause in the random scalar field $\phi$, say through a scaling $\phi^*=\Lambda\phi$, then $f_2$ has to transform as a quadratic form, i.e., in the just chosen scaling then particularly as $f_2^*=\Lambda^{-2}f_2$, while $f_1$ as the single form $f_1^*=\Lambda^{-1}f_1$. But such a transformation is only possible if $m=6$ --- but when choosing $m=6$, there is the problem again that no inverse energy cascade will develop, as was shown in \cite{Falkovich10}.

This concludes our proof that the conformal invariance as proposed in \cite{Grebenev21} violates the classical principle of cause and effect for $m\neq6$, and therefore does not map solutions to solutions as incorrectly claimed.

\appendix

\section{The infinite chain of moment equations and its invariances}

Although for the conformal invariance in \cite{Grebenev21} only the first (lowest order) statistical equation was considered, as we also did when presenting the invariance \eqref{211102:0004}, we now want to extend this invariance analysis to the entire infinite chain of moment equations. As this chain reads \cite{Gihman72}
\begin{equation}
\hat{\alpha}\langle\hat{\phi}^n\rangle-\rho\langle\hat{\phi}^{n+2}\rangle+(n-1)\,\hat{Q}\,\langle\hat{\phi}^{n-2}\rangle=0,\quad n\geq 1,
\label{211107:1515}
\end{equation}
we can exploit the fact that on each level $n$ up to infinite order the next higher-order moment $\langle\hat{\phi}^{n+2}\rangle$ in the above equation is an unclosed term, a degree of freedom that allows us to construct any desirable invariance. For example, the arbitrary choice of the invariance \eqref{211102:0004} in the two lowest-order moments then iteratively transfers to all higher orders as
\begin{equation}
\left.
\begin{aligned}
&\langle\hat{\phi}^{n+2}\rangle^*=e^{a_s}\langle\hat{\phi}^{n+2}\rangle-T_{n+2}/\rho,\quad n\geq 3,\\[0.5em]
&\text{with}\;\; T_{n+2}=\hat{\alpha}\,\Big(e^{a_s}\langle\hat{\phi}^{n}\rangle-\langle\hat{\phi}^{{n}}\rangle^*\Big)
+(n-1)\,\hat{Q}^*\,\Big(\langle\hat{\phi}^{n-2}\rangle-\langle\hat{\phi}^{{n-2}}\rangle^*\Big),
\end{aligned}
~~~\right\}
\end{equation}
leaving invariant then to all orders the chain of equations \eqref{211107:1515} with the initialization \eqref{211102:0004} for $n\leq2$.\linebreak[4]
But also any other desirable invariance can be designed for the lowest-order moments and then prolonged to all higher orders. For example, the alternative choice
\begin{equation}
\left.
\begin{aligned}
\!\!\!&\rho^*=e^{-a_s}\rho,\;\;\: \hat{\alpha}^*=\hat{\alpha},\;\;\: \hat{Q}^*=Q,\;\;\: \langle\hat{\phi}^2\rangle^*=\langle\hat{\phi}^2\rangle,\;\;\:
\langle\hat{\phi}^{n+2}\rangle^*=e^{a_s}\Big(\langle\hat{\phi}^{n+2}\rangle-T^\prime_{n+2}/\rho\Big),\;\;\: n\geq 2,\\[0.5em]
\!\!\!&\text{with}\;\; T^\prime_{n+2}=\hat{\alpha}\,\Big(\langle\hat{\phi}^{n}\rangle-\langle\hat{\phi}^{{n}}\rangle^*\Big)
+(n-1)\,\hat{Q}\,\Big(\langle\hat{\phi}^{n-2}\rangle-\langle\hat{\phi}^{{n-2}}\rangle^*\Big),
\end{aligned}
~~\right\}
\label{211107:1714}
\end{equation}
is another example of a so-called ``intermittency symmetry" \cite{Oberlack14,Oberlack17}, which too leaves invariant \eqref{211107:1515}, but which, as explained in Sec.$\,$\ref{Sec4}, is again not realizable and therefore nonphysical, since it again simply violates the principle of cause and effect of the defining stochastic equation \eqref{211030:1406}.

Worthwhile to note is that in the limit of zero stochastic forcing $Q\to 0$ (careful, not exact zero~forcing $Q=0$, but zero forcing in the limit), the infinite hierarchy of moment equations \eqref{211107:1515} reduces to
the asymptotic form
\begin{equation}
\hat{\alpha}\langle\hat{\phi}^n\rangle-\rho\langle\hat{\phi}^{n+2}\rangle=0,\quad n\geq 1,
\label{211108:0949}
\end{equation}
with the solution
\begin{equation}
\langle\hat{\phi}^n\rangle=
\begin{cases}
\:(\hat{\alpha}/\rho)^{n/2},\;\; \text{$n$ even},\\[0.5em]
\:0,\;\; \text{$n$ odd},
\end{cases}
\label{211108:1041}
\end{equation}
being essentially the asymptotic average of the deterministic solution ($Q=0$), which consists of two stable equilibrium solutions $\hat{\phi}=\pm(\hat{\alpha}/\rho)^{1/2}$, where the positive or negative solution depends on the chosen sign of the initial condition, separated by the unstable zero-solution $\hat{\phi}=0$ for the exact initial condition $\hat{\phi}(0)=0$.
While the above statistical solution \eqref{211108:1041} can be straightforwardly obtained through the steady-state PDF \eqref{211108:1044} in the limit $Q\to 0$, it's more intricate to obtain it numerically when solving the stochastic equation \eqref{211030:1406} directly. It's expedient not to take the time average but rather the ensemble average for a set of initial conditions around the unstable equilibrium point at~$\hat{\phi}=0$ and
then letting the initial point range go to zero as $Q\to 0$.

The infinite chain of moment equations \eqref{211108:0949} allows now for an ``intermittency symmetry" of the following ``pure" form
\begin{equation}
\langle\hat{\phi}^n\rangle^*=e^{a_s}\langle\hat{\phi}^n\rangle,\quad
\hat{\alpha}^*=\hat{\alpha},\quad
\rho^*=\rho,\quad n\geq 1,
\label{211108:1549}
\end{equation}
being fully identical now to the ``symmetry" considered in \cite{Oberlack14,Oberlack17}, because it arises from the statistical equations of Navier-Stokes turbulence under the same conditions as \eqref{211108:1549} results from \eqref{211108:0949}, namely just driven by the instability of the system and not by a stochastic force.

Although \eqref{211108:1549} leaves invariant \eqref{211108:0949} to all orders again, it's nevertheless a nonrealizable and thus nonphysical invariance again, for the very same reason as before: it violates again the principle of cause and effect. No transformation of any kind in the random field variable exists, whether as an invariance or not, such that~\eqref{211108:1549} will result. In other words, for the occurring change in the averaged field variables in~\eqref{211108:1549} there is no origin for it in its defining stochastic equation \eqref{211030:1406} (for $Q\to 0$). Simply put, there is no transformation, say of the following general point-form
\begin{equation}
t^*=t^*(t,\hat{\phi}),\quad \hat{\phi}^*=\hat{\phi}^*(t,\hat{\phi}),\quad
\hat{\alpha}^*=\hat{\alpha},\quad
\rho^*=\rho,
\label{211108:1638}
\end{equation}
such that when transforming the stochastic equation according to \eqref{211108:1638} to a new equation in the variables~$(t^*,\hat{\phi}^*)$, and then generating from it the new statistical solutions, that it will result
to~\eqref{211108:1549}.\linebreak[4] Or, said differently, there is no transformation in the random field variable \eqref{211108:1638}, whether as an invariance or not, such that when averaging this transformation that it will result to \eqref{211108:1549}. The~problem lies trivially in the fact that all moments scale identically to all orders with the same scaling exponent, a~statistical scaling feature that is trivially nonphysical \cite{Frewer15,Frewer16,Frewer17,Frewer14}. Hence, obviously, \eqref{211108:1549} does not map solutions to solutions. For example, if one maps the statistical solution \eqref{211108:1041}, one obtains the new transformed moments
\begin{equation}
\langle\hat{\phi}^n\rangle^*=
\begin{cases}
\:e^{a_s}(\hat{\alpha}/\rho)^{n/2},\;\; \text{$n$ even},\\[0.5em]
\:0,\;\; \text{$n$ odd},
\end{cases}
\label{211108:1704}
\end{equation}
which evidently cannot be generated as solutions from the stochastic equation \eqref{211030:1406} (for $Q\to 0$), or any other one, no matter which approach one takes.

For a more general discussion on equivalence transformations in turbulence and their realizability, see for example the final remarks R4 or R5 in \cite{Frewer18,Frewer21.1} and the references therein.

\nocite{apsrev42Control}
\bibliographystyle{apsrev4-2}
\bibliography{References}

\end{document}